\documentclass[aps, prb, twocolumn, showpacs, showkeys, nofootinbib]{revtex4}
\usepackage{latexsym}
\usepackage{graphicx}
\usepackage{amsmath}
\usepackage{graphicx}
\usepackage{amssymb}
\usepackage{mathrsfs}
\usepackage{bbm}
\usepackage{bm}
\usepackage{multirow}
\usepackage{relsize}
\usepackage{lineno}
\usepackage[sort&compress]{natbib}
\usepackage{dcolumn}%

\begin{document}


\newcommand{\dd}{d}
\newcommand{\pd}{\partial}
\newcommand{\myU}{\mathcal{U}}
\newcommand{\myr}{q}
\newcommand{\Urho}{U_{\rho}}
\newcommand{\myalpha}{\alpha_*}
\newcommand{\bd}[1]{\bm{#1}}
\newcommand{\Eq}[1]{Eq.\ (\ref{#1})}
\newcommand{\Eqn}[1]{Eq.\ (\ref{#1})}
\newcommand{\Eqns}[1]{Eqs.\ (\ref{#1})}
\newcommand{\myS}{Sec.\ }
\newcommand{\figsize}{8.6cm}
\newcommand{\Figref}[1]{Fig.~\ref{#1}}
\newcommand{\Figrefs}[1]{Figs.~\ref{#1}}

\preprint{Geometric Phase of Asymmetric Top}

\title{Analytic formula for the Geometric Phase of an Asymmetric Top}

\author{Nicholas A. Mecholsky}
 \email{nmech@vsl.cua.edu}
 \thanks{Corresponding Author}

\affiliation{
Vitreous State Laboratory\\
The Catholic University of America\\
Washington, DC 20064}

\date{1/21/2019}

\begin{abstract}
The motion of a handle spinning in space has an odd behavior.  It seems to unexpectedly flip back and forth in a periodic manner as seen in a popular YouTube video.\cite{thandle}  As an asymmetrical top, its motion is completely described by the Euler equations and the equations of motion have been known for more than a century.  However, recent concepts of the geometric phase have allowed a new perspective on this classical problem.  Here we explicitly use the equations of motion to find a closed form expression for total phase and hence the geometric phase of the force-free asymmetric top and explore some consequences of this formula with the particular example of the spinning handle for demonstration purposes.  As one of the simplest dynamical systems, the asymmetric top should be a canonical example to explore the classical analog of Berry phase.
\end{abstract}


\maketitle

\section{Introduction}
The motion of spinning objects is ubiquitous.  From a seemingly mundane and classical motion, new mysteries are still being debated and surprising features are revealed.\cite{moffatt2000euler, rollingrings, berry2010slow}  One important observation is the identification of a Berry phase of rotating classical objects.\cite{berry1988classical, montgomery1991much, hannay1985angle} Geometric phases have become popular recently due to their applications to material properties (see Ref.\ (\onlinecite{xiao2010berry}) for a recent review) and other theoretical and calculational advantages.\cite{resta2000manifestations, mead1992geometric}  Examples have relied on Berry's original paper\cite{berry1984} describing the quantum mechanical effect and relegating the classical version to the term `classical analog'.  The classical analog of Berry phase is the Hannay angle and was developed in the mid 1980s by Hannay and others, inspired by Berry's original work.\cite{hannay1985angle, montgomery1991much}  The canonical example of the Hannay angle usually involves Foucault's pendulum.\cite{anandan1992geometric, berry1988geometric}  However, even in the force-free spinning of a top, the Hannay angle is apparent, yet not typically discussed in elementary treatments of the subject.  Evidently Jacobi was the first to write down the exact analytic expressions for the motion of the asymmetric top in the 19th century.  In Landau and Lifshitz's Mechanics,\cite{LL} it was already appreciated that the motion of the top was not perfectly periodic in time.  This was identified by Landau and Lifshitz when they stated ``this incommensurability has the result that the top does not at any time return to its original position." (see Ref.\ (\onlinecite{LL}), \S 37, page 120)  However, this observation was not investigated further in that text and the connection to the Hannay angle was not included in their work which would have predated the original work by Hannay\cite{hannay1985angle} and Berry.\cite{berry1984}

Geometric phases (Hannay angles) are not typically discussed in undergraduate courses. The basic features and calculations can be seen in several references.\cite{hannay1985angle, lawson2016spacewalks, gil2010mechanical, berry1988classical, jose2000classical, marsden1992lectures} The geometric phases of several real-world examples are calculated in Ref.\ (\onlinecite{lawson2016spacewalks}). As a simple example, consider a classical system that has periodic motion.  An example of this is the asymmetric top in force-free motion in space.  Some of the dynamic variables are periodic (like the angular momentum and angular velocity vectors).  However other variables (like the Euler angles) are not periodic.  In one cycle of the periodic variables, the system almost returns to its original state except for the variables that are not periodic.  Typically this can be viewed as an angular difference from the initial state.  This extra phase will be referred to as the `total phase' in this paper.  The total phase is a combination of two parts.  One is the dynamic part that depends on the time dynamics and the other is the geometric part, called the Hannay Angle,\cite{robbins2016hannay} that is independent of time and hence `geometric'.  Alternatively, we may consider a periodic system whose Hamiltonian is forced to change slowly due to some external influence.  An example of this could be the Foucault pendulum.\cite{berry1988geometric, hart1987simple}  If the system is then returned to the original Hamiltonian, there is an extra phase due to this forcing.  One might call these two examples the `passive' and `active' cases.

In 1991, Richard Montgomery, and around the same time, Mark Levi, published papers\cite{montgomery1991much, levi1993geometric} that derived a formula for computing the total phase of rigid bodies
\begin{equation}\label{eqn:montgomery}
\Delta \alpha = \frac{2 E T}{M} - \Omega,
\end{equation}
where $T$ is the period of the angular velocity vector in space (here we compute it from \Eqn{eqn:TK}) and $\Omega$ is the signed solid area swept out by the angular momentum vector.  The dynamic part $\left( \dfrac{2 E T}{M} \right)$ is the integrated angle of the angular velocity projected onto the angular momentum vector.  The geometric part is a fraction of a unit sphere swept out by the path of the angular momentum vector in the body frame (using the right hand rule to assign a sign).

Even though much work has been done in this area in the past few decades (see for example Ref.\ \onlinecite{bates2005rotation}), the connection and explicit formula for the total phase of an asymmetric top was never published. Part of the goal of this paper is to take the last step and cast the exact expressions for the motion of the asymmetric top in the framework of the total phase.  A closed form expression for the total phase will be presented and explored.  We use the example of the T-shaped handle (seen flipping around on the International Space Station in a recent YouTube video\cite{thandle}) to demonstrate some of these expressions, however the formulas are valid for a general asymmetric top.  In some ways, this is the most elementary example of total phase available in classical physics.  Even in this simple system, a wide variety of questions and lines of investigation are possible.  To that end, an exploration and some observations of the formula are in order.

The paper begins by reviewing the dynamics of the asymmetric top.  The angular velocity and Euler angles are determined as functions of time.  From these expressions, the total phase is determined directly and compared to Montgomery's formula.  Finally, some observations about the total phase are made and some conclusions are discussed.

\section{Dynamics and Total Phase}
\begin{figure}[!hb]
	\begin{center}
 	\includegraphics[width=\figsize]{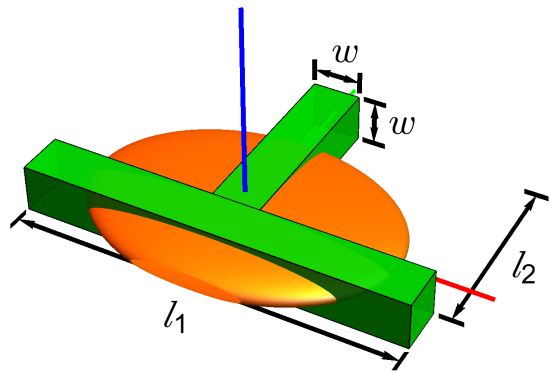}\\
      \caption{\label{fig:fig1}  Two equivalent rigid bodies.  A T-shaped handle (green) with $l_1 = 8$, $l_2 = 4$, and $w = 1$ and the equivalent solid ellipsoid that has the identical principal moments of inertia given by \Eqns{eqn:ellipsoid}, with semi-axes $(a, b, c) = (4.23281, 3.07318, 0.645497)$.  The ellipsoid is shrunk by 1/4 for presentation.}
	\end{center}
\end{figure}

As an example for visualization, and without loss of generality, we may use a T-shaped handle to explore our dynamics such as the dynamics seen in the video.\cite{thandle}  For any three physical moments of inertia there is a corresponding T-shaped handle and thus we are losing no generality.  Consider a handle of uniform density and unit mass with a cross piece of length $l_1$, a shaft of length $l_2$, and a square cross-sectional width of $w$ (\Figref{fig:fig1}).
%
%
%
%
%
%

The total moments of inertia can be computed using the parallel axis theorem with individual square prisms.  This will give a moment of inertia tensor with diagonal elements
%
%
\begin{subequations}\label{eqn:IsforT}
\begin{align}
&I_{1} = \frac{6 l_1 l_2 w^2+2 l_1^2 w^2+l_2^2 w \left(6 l_1+w\right)+l_2^4+4 l_1 l_2^3}{12 \left(l_1+l_2\right)^2 } \\
&I_{2} = \frac{\left(l_1+2 l_2\right) w^2+l_1^3}{12 \left(l_1+l_2\right)} \\
&I_{3} = \nonumber\\
&\frac{l_1 l_2 \left(l_1^2+5 w^2 + 4 l_2^2\right)+l_1^2 \left(l_1^2+w^2\right)+l_2^2 w \left(6 l_1+w\right)+l_2^4}{12 \left(l_1+l_2\right)^2},
\end{align}
\end{subequations}
%
%
where the coordinate axes have been chosen to have a diagonal moment of inertia tensor.  An ellipsoid (with semi-axes lengths $a$, $b$, and $c$) with the same moment of inertia is given by 
\begin{subequations}\label{eqn:ellipsoid}
\begin{align}
a &= \frac{\sqrt{\frac{5}{3}} \sqrt{l_2 w^2+l_1^3}}{2 \sqrt{l_1+l_2}} \\
b &= \frac{\sqrt{\frac{5}{3}} \sqrt{2 l_1 l_2 \left(3 l_2 w+2 l_2^2+2
   w^2\right)+l_1^2 w^2+l_2^4}}{2 \sqrt{\left(l_1+l_2\right){}^2}} \\
c &= \frac{1}{2}
   \sqrt{\frac{5}{3}} w,
\end{align}
\end{subequations}
shown superimposed in \Figref{fig:fig1}.  Although this is correct for all $l_1$, $l_2$, and $w$, here we wish $l_1$ large enough\footnote{Otherwise a relabeling of $a$, $b$, and $c$ is needed so that the moments of inertia are correctly ordered.} with respect to $l_2$ (and $w$ small enough) so that $a>b>c$.

It will turn out that the only important parameters for the description of the body are those that determine the direction that the moment of inertia points in the moment of inertia space (the polar and azimuthal angles relative to the $I_1, I_2,$ and $I_3$ axes).  Thus the absolute magnitude of the vector of the eigenvalues of the moment of inertia is not important.  We will find it useful to use a moment of inertia that is parameterized by two variables, $b$ and $c$, the semi-axes of a solid ellipsoid.  The third semi-axis of this ellipsoid is determined such that the moment of inertia vector (composed from the eigenvalues) has unit magnitude.  The moment of inertia for this particular choice is given by,
\begin{subequations}\label{eqn:isforellipsoid}
\begin{align}
I_{1} &= b^2+c^2 \\
I_{2} &= \frac{1}{2} \left(-b^2+\sqrt{2-3 b^4-2 b^2 c^2-3 c^4}+c^2\right) \\
I_{3} &= \frac{1}{2} \left(b^2+\sqrt{2-3 b^4-2 b^2 c^2-3 c^4}-c^2\right)
\end{align}
\end{subequations}
with the third semi-axis of the ellipsoid given by 
\begin{equation}
a = \frac{\sqrt{\sqrt{2 - 3 b^4 - 2 b^2 c^2 - 3 c^4} - b^2 - c^2}}{\sqrt{2}}
\end{equation}
This particular choice of semi-axes of the ellipsoid is always chosen to be labeled so that $I_1 < I_2 < I_3$. 
However, as long as $a>b>c$ in the ellipsoid, this will always be the case.

In the next section we take a small digression to identify the full space of moments of inertia.

\subsection{Space of Possible Moments of Inertia}
Not all triplets $(I_1, I_2, I_3)$ of positive numbers are valid moments of inertia.  To have a valid (physical) moment of inertia we must satisfy the following relations,
\begin{align}
I_1 + I_2 &\geq I_3 \\
I_2 + I_3 &\geq I_1 \\
I_3 + I_1 &\geq I_2.
\end{align}
these may be referred to as the inertial inequalities.

\begin{figure}[!hb]
	\begin{center}
 	\includegraphics[width=\figsize]{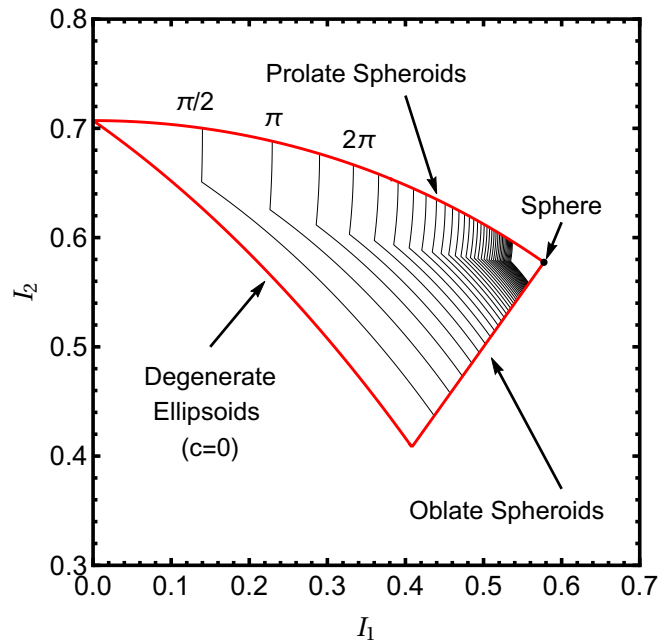}\\
      \caption{\label{fig:fig2}  Full parameter space of asymmetric rigid bodies of fixed moment of inertia magnitude equal to 1.  $I_1$ and $I_2$ are two of the principal moments of inertia labeled so that $I_1 < I_2 < I_3$, and the magnitude of the moment of inertia vector is unity. The red curves are degenerate asymmetric rigid bodies, here modeled as solid ellipsoids: Prolate spheroids (rod-like shapes), spheres, oblate spheroids (coin-like shapes), and degenerate ellipsoids (lines and disks).  The equations for these curves are given in the text. The contours are curves of constant minimum values for the total phase.  Given a body (parametrized by two of the three values of the moment of inertia), any initial condition will lead to at least a total phase of the given value indicated by the contour values.}
	\end{center}
\end{figure}

Without loss of generality (we may relabel axes in the object if necessary) we may further restrict ourselves to the region, $I_1 < I_2 < I_3$.  For labeling purposes, consider only the moments of inertia where $I_1^2+I_2^2+I_3^2 = 1$. In this case (we will see that these are the only cases we need consider) we can show the region of possible physical moments of inertia as those moments of inertia inside the red boundary in the $I_1 - I_2$ plane in \Figref{fig:fig2}.  The curves that bound this region are the curve of all prolate spheroids $\left( I_2 = \sqrt{\dfrac{1-I_1^2}{2}} \right)$, the curve of all oblate spheroids $\left( I_2 = I_1 \right)$, and a curve of degenerate ellipsoids where $c=0$, $I_2 = \dfrac{1}{2} \left( \sqrt{2 - 3 I_1^2} - I_1\right)$.  Thus picking $0 < c < b < 1$ fixes a particular asymmetric top.  For the rest of the paper, we assume a rigid object with moments of inertia fixed.
%
%
%
%
%
%
%

\subsection{Exact Dynamics of the Asymmetric Top}
Suppose we decide to spin this asymmetric top around some axis with some initial angular velocity and without any external torques.  Let us say that in the body frame of reference (determined by the principle moments of inertia $I_1 < I_2 < I_3$ in a right handed coordinate system), the initial angular velocity is given by
\begin{equation}
\bd{\omega}_0 = (\omega_{01},\omega_{02},\omega_{03}).
\end{equation}
We will show below that the total phase will not depend on the magnitude of $\bd{\omega}_0$.  Thus we shall take $| \bd{\omega}_0 | = 1$.

\begin{figure}[!hb]
	\begin{center}
 	\includegraphics[width=\figsize]{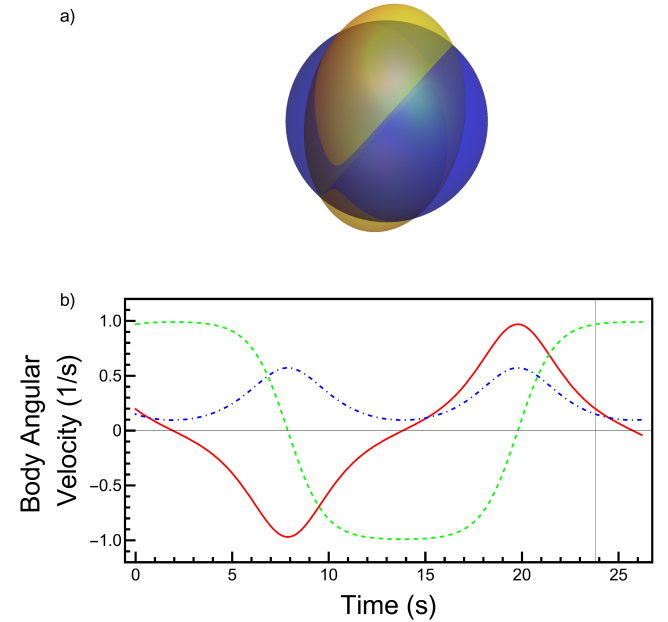}\\
      \caption{\label{fig:fig3}  a) Intersection in angular momentum space of two constants of the motion: the constant moment of inertia sphere (blue) and the energy ellipsoid (yellow). b) Representative motion of the body angular velocity in time.  The solid red (dashed green, dot-dashed blue) curve plots the value of the angular velocity component along the body $x$ ($y$, $z$) axis which corresponds to the smallest (intermediate, largest) moment of inertia.  We choose the total mass (from a uniform density) of the T-handle in \Figref{fig:fig1} to be such that the moment of inertia has unit magnitude.  The initial angular velocity has unit magnitude, and the direction is given by a polar angle (measured from the body $z$ axis), $\theta_{\omega} = \pi/2 -  0.15$, and an azimuthal angle, $\phi_{\omega} = \pi/2 - 0.2$, measured from the body $x$ axis.  This sets the moment of inertia and total rotational energy.  Those, along with the moments of inertia, are: $E = 0.264805 \, J$, $M = 0.533252 \, J s $, $I_1 = 0.286827 \, J s^2$, $I_2 = 0.533256 \, J s^2$, and $I_3 = 0.795844 \, J s^2$.  The period of the motion is 23.8181 s, marked with a gray vertical line in (b).}
	\end{center}
\end{figure}

The time dynamics of the angular velocity is then determined by these initial conditions.  In torque-free motion, there are two constants of motion for the asymmetric top.  One is the total angular momentum of the top, given by
\begin{equation}\label{eqn:J}
M = | \bd{I} \cdot \bd{\omega}_0 | = \sqrt{I_1^2 \omega_{01}^2+ I_2^2 \omega_{02}^2 + I_3^2 \omega_{03}^2},
\end{equation}
and the total energy of rotation (assume the center of mass velocity of the asymmetric top is 0), given by
\begin{equation}\label{eqn:E}
E = \frac{1}{2} \bd{\omega}_0 \cdot \bd{I} \cdot \bd{\omega}_0 = \frac{1}{2} \left( I_1 \omega_{01}^2 + I_2 \omega_{02}^2 + I_3 \omega_{03}^2 \right).
\end{equation}
The motion of the angular momentum vector in space coordinates is, of course, fixed since no torque is applied to change the angular momentum.  In the body coordinates, the angular momentum vector executes periodic motion, where the origin is fixed and the terminus sweeps out a curve in the body frame.  The curve that the angular momentum sweeps out in space is constrained to be the intersection of these two constants of motion.  This intersection curve is called the polhode.  If we use $M_i = I_i \omega_{0i}$, then we may rewrite \Eqns{eqn:J} and (\ref{eqn:E}) slightly to give 
\begin{align}\label{eqn:MM}
M^2 &= M_1^2 + M_2^2 + M_3^2\\ \label{eqn:EM}
E &= \frac{1}{2} \left( \frac{M_1^2}{I_1} + \frac{M_2^2}{I_2} + \frac{M_3^2}{I_3} \right)
\end{align}
which are the equivalent to Eqns.\ 5.45 and 5.46, p.\ 203-204 of Ref.\ (\onlinecite{goldstein}), or Eqns.\ 37.3 and 37.4, p.\ 116 \S 37 of Ref.\ (\onlinecite{LL}).  This intersection can be seen in \Figref{fig:fig3} (a).
%
%
%
%
%
%

Equations (\ref{eqn:MM}) and (\ref{eqn:EM}) represent a sphere of constant magnitude (the angular momentum magnitude is fixed), and an intersecting ellipsoid that is the surface of constant energy in angular momentum coordinates (a space where the coordinates are the projections of the angular momentum along the body's principle moments of inertia) called the Binet ellipsoid.\cite{goldstein}

The two constant surfaces must intersect for a physical object and this means that the constant angular momentum sphere is smaller than largest Binet ellipsoid axis and bigger than smallest Binet ellipsoid axis.  This gives the condition
\begin{equation}
2 E I_1 \leq M^2 \leq 2 E I_3
\end{equation}
This will always happen since the alternative is for no intersection of the two constant surfaces, and thus no physical dynamics.  There are two possible types of curves on the angular momentum sphere and the Binet ellipsoid (not including degenerate cases).  To avoid degenerate cases (discussed elsewhere for example Ref.\ \onlinecite{LL} \S 37, p.\ 119 or Ref.\ \onlinecite{goldstein}) we shall always assume the strict inequalities.  We shall also take $M^2 > 2 E I_2$ for the equations below, but for the case $M^2 < 2 E I_2$, the formulas require $1 \leftrightarrow 3$ in the indices.

The dynamics of the angular velocity in the body frame evolve according to the Euler equations
\begin{subequations}\label{eqn:Euler}
\begin{align}
I_1 \dot{\omega}_1(t) &= (I_2 - I_3) \omega_2(t) \omega_3(t), \\
I_2 \dot{\omega}_2(t) &= (I_3 - I_1) \omega_3(t) \omega_1(t), \\
I_3 \dot{\omega}_3(t) &= (I_1 - I_2) \omega_1(t) \omega_2(t).
\end{align}
\end{subequations}

This may be solved exactly in terms of Jacobian elliptic functions (Eq.\ 37.10 of Ref.\ \onlinecite{LL}, \S 37, p.\ 118).  The motion of the angular velocity vector is perfectly periodic in time with period
\begin{subequations}\label{eqn:TK}
\begin{align}
T(E, M, I_1, I_2, I_3) &= \frac{4 \, \mathlarger{\textrm{K}}( k(E, M, I_1, I_2, I_3)^2 )}{\sqrt{\frac{(I_3 - I_2) (M^2 - 2 E I_1)}{I_1 I_2 I_3}}} \quad \textrm{with} \\
k(E, M, I_1, I_2, I_3) &= \sqrt{\frac{(I_2 - I_1) (2 E I_3 - M^2)}{(I_3 - I_2) (M^2 - 2 E I_1)}},
\end{align}
\end{subequations}
where $\mathlarger{\textrm{K}}(m)$ is the Complete Elliptic Integral of the First Kind, namely,
\begin{equation}\label{eqn:K}
\mathlarger{\textrm{K}}(m) = \int_0^{\pi/2} \frac{\dd \theta}{\sqrt{1 - m \sin^2 \theta}}.
\end{equation}
In some notations, the argument of the Complete Elliptic Integral of the First Kind, $\mathlarger{\textrm{K}}$, is $\sqrt{m}$ and not $m$.  We choose the notation in \Eqn{eqn:K}, but care must be taken here.  Note that when $M^2 < 2 E I_2$, replace $1 \leftrightarrow 3$ in the quantities above.  This has the effect of making $T$ always real and $0<k<1$.

Given the initial angular velocity vector $\bd{\omega}_{0}$ in the body frame with polar angle $\theta_{\bd{\omega}}$ from the body $z$ axis and azimuthal angle $\phi_{\bd{\omega}}$ from the body $x$ axis, the initial angular momentum ($\bd{M}_0 = \bd{I} \cdot \bd{\omega}_0$) is not collinear with the angular velocity (since $I_1 \neq I_2 \neq I_3$).  The initial angular momentum is fixed in the space frame and we may define the $z$ axis of the space coordinates to be this angular momentum vector.  We define $\theta_{\omega}$ as the angle that $\bd{\omega}_{0}$ makes with this axis. 

Since the angular velocity vector is explicitly determined in time (See the Appendix \Eqns{eqn:Omegas} and Ref.\ \onlinecite{LL} Eq.\ (37.10)), we may further express the Euler angles as explicit functions of time.  Here we choose the Euler angle convention of both Refs.\ (\onlinecite{goldstein}) and (\onlinecite{LL}) (Figure 4.7, p.\ 152 and Fig.\ 47 \S 35, respectively).  This describes a sequence of intrinsic elemental rotations about the body $z$-$x'$-$z''$ axes. In this convention, two of the Euler angles ($\theta$ and $\psi$) are given algebraically in terms of the angular velocity functions, and the third ($\phi$) is given as a first-order differential equation in time. Reference \onlinecite{LL} claimed that this may be further integrated to give $\phi$ in terms of theta functions, referencing Chapter 6 of Ref.\ \onlinecite{whittaker1988treatise} and arriving at Eq.\ 37.18-37.20.  However we were not able to verify this solution due to differences in theta function notation in many sources.  However, a straightforward solution is provided in the Appendix for all Euler angles in \Eqns{eqn:eulerangles} and (\ref{eqn:fullphi}).  Briefly, the integration for $\phi$ may be carried out directly to give Incomplete Elliptic Integrals of the Third Kind. Please see the Appendix for details.

\subsection{Formula for Total Phase}
Even though the equation for the angular velocity is exactly periodic in time, the orientation and angles after a period are not periodic and will not in general return to the same orientation.  This angular mismatch after a period of the angular velocity will be called the total phase (or the same quantity mod 2$\pi$). The term `geometric phase' (or Hannay angle) is used to refer to just the geometric part of this phase.\cite{montgomery1991much, robbins2016hannay, zwanziger1990berry}  We reserve `Berry phase' as the quantum mechanical version of the geometric phase where a wave function is explicitly transported in parameter space.  In contrast, we are investigating the natural (so to speak) phase associated with force-free evolution.

In computing the total phase, we could use Montgomery's formula,\cite{montgomery1991much} \Eqn{eqn:montgomery},
\begin{equation}
\Delta \alpha = \frac{2 E T}{M} - \Omega,
\end{equation}
where $T$ is the period from \Eqn{eqn:TK} and $\Omega$ is the signed solid area swept out by the angular momentum vector.  The dynamic part $\left( \dfrac{2 E T}{M} \right)$ is the integrated angle of the angular velocity projected onto the angular momentum vector over one period.  We may identify the total phase from the exact difference of the Euler angles over one period. The geometric part is the fraction of a unit sphere swept out by the angular momentum vector over one period (using the right hand rule to assign a sign). This might be a difficult computation to do since it is a geometric surface area enclosed by the polhode.  It may still be done numerically.

However considering that we have exact expressions for the Euler angles (\Eqns{eqn:eulerangles} and (\ref{eqn:fullphi})), we may evaluate the total phase directly.  The angular momentum vector after one period will be identical to its starting value.\cite{montgomery1991much} Thus the angular difference around this axis is the total phase.  Consider that in some cases (when $M^2 > 2 E I_2$), this may involve $\psi$ as well $\phi$, but not $\theta$.  In this case, $\psi$ always changes by $2 \pi$.  When $M^2 < 2 E I_2$, $\psi$ is perfectly periodic and thus the angular change in $\phi$ is zero.  If we assume that the Euler angle $\phi$ at time $t = 0$ is 0, then we have the change in angle, $\Delta \alpha$, over one period is
\begin{equation}\label{eqn:alphaproto}
\Delta \alpha =\begin{cases} 
      \phi(T) - 2 \pi & M^2 > 2 E I_2 \\
      \phi(T) & M^2 < 2 E I_2 
   \end{cases}.
\end{equation}
Plugging in the expressions for $\phi$ in the Appendix (\Eqns{eqn:fullphi}), \Eqn{eqn:alphaproto} reduces to 
\begin{widetext}
\begin{equation}\label{eqn:GeometricPhase}
\Delta \alpha = \begin{cases}
 \frac{M T}{I_3} + \frac{4 M \left( I_3 - I_1 \right)}{I_1 I_3} \frac{\mathlarger{\Pi}\!\left(- \frac{I_3 \left(I_2-I_1\right)}{I_1 \left(I_3-I_2\right)} \big|  k^2 \right)}{\sqrt{\frac{(I_3 - I_2) (M^2 - 2 E I_1)}{I_1 I_2 I_3}}} -  2 \pi & M^2 > 2 E I_2 \\ 
\frac{M T}{I_3} + \frac{4 M \left( I_3 - I_1 \right)}{I_1 I_3} \frac{\mathlarger{\Pi}\!\left(-\frac{I_3 \left(M^2-2 E I_1\right)}{I_1 \left(2 E I_3 - M^2\right)} \big| k^2 \right)}{\sqrt{\frac{(I_2 - I_1) (2 E I_3- M^2)}{I_1 I_2 I_3}}} & M^2 < 2 E I_2
\end{cases}
\end{equation}
\end{widetext}
where $\mathlarger{\Pi}\!\left( n | m \right)$ is the Complete Elliptic Integral of the Third Kind defined by
\begin{equation}\label{eqn:Pi}
\mathlarger{\Pi}\!\left( n \big| m \right) = \int_0^{\pi/2} \frac{\dd \theta}{(1 - n \sin^2 \theta) (1 - m \sin^2 \theta)^{1/2}},
\end{equation}
where $k$ is defined in \Eqn{eqn:TK} (b).

It is worthwhile to note that the total phase formula in \Eqn{eqn:GeometricPhase} seems to depend on $E$, $M$, $I_1$, $I_2$, and $I_3$. However it can be shown that it does not depend on the magnitude of the moment of inertia ($I^2 = I_1^2+I_2^2+I_3^2$) nor on the magnitude of the initial angular velocity.

To see the independence of the magnitude $I$, note that the energy, \Eqn{eqn:E}, is homogeneous of order 1 in $I$.  The angular momentum, \Eqn{eqn:J}, is also homogeneous of order 1 in $I$.  The functions $k$ and $T$ of \Eqns{eqn:TK} are therefore homogeneous of order 0 in $I$.  It then follows that \Eqn{eqn:GeometricPhase} is independent of the magnitude of the moment of inertia.

For the independence of the magnitude of $\bd{\omega}_{0}$, let us consider the initial polar and azimuthal angles of the angular velocity ($\bd{\omega}_{0}$) with respect to the body axes ($\theta_{\bd{\omega}}$ and $\phi_{\bd{\omega}}$) fixed for the moment.  Note that in this case, the angular momentum is homogeneous of order 1 in $|\bd{\omega}_{0}|$, and the energy is homogeneous of order 2 in $|\bd{\omega}_{0}|$.  This means that $k$ is homogeneous of order 0 in $|\bd{\omega}_{0}|$ and $T$ is homogeneous of order -1 in $|\bd{\omega}_{0}|$. Then looking at the total phase formula, \Eqn{eqn:GeometricPhase}, this means that the total phase is independent from the magnitude of the initial angular velocity, $|\bd{\omega}_{0}|$. 

Considering that the total phase formula is itself a constant of motion (it is composed of constants of motion), the initial angles, $\theta_{\bd{\omega}}$ and $\phi_{\bd{\omega}}$, which set the constants of the motion ($E$ and $M$) along with the direction of the moment of inertia (or simply $I_1$ and $I_2$ along with the constraint that $I_1^2+I_2^2+I_3^2=1$) are the only parameters needed to determine the total phase.  Thus every point along the closed curve on the Binet ellipsoid has the same total phase since every point has the same quantities of $E$, $M$, $I_1$, $I_2$, and $I_3$.

\begin{figure}[!hb]
	\begin{center}
 	\includegraphics[width=\figsize]{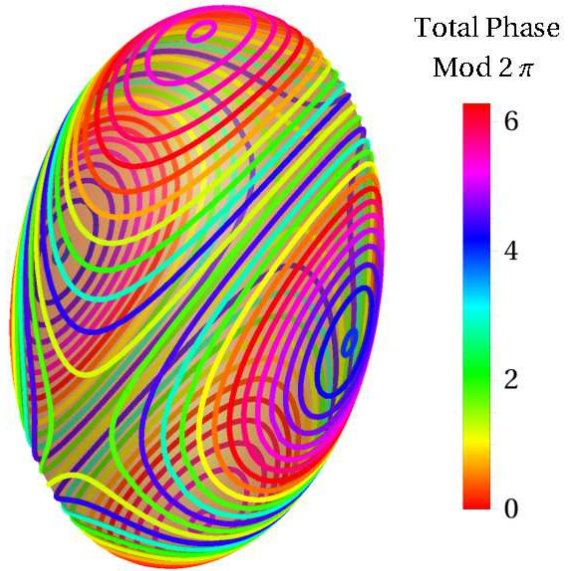}\\
      \caption{\label{fig:fig4} Total phase (mod 2$\pi$) of the body given in \Figref{fig:fig1} for an asymmetric top with ($I_1$, $I_2$, $I_3$) = ($0.286827 \, J s^2$, $0.533256 \, J s^2$, $0.795844 \, J s^2$).  For each initial angle, a given curve on the Binet ellipsoid is represented.  The Binet ellipsoid is the ellipsoid of constant energy in the angular momentum space.  Given an initial angular momentum, the resulting dynamics traces out a closed curve on this surface.  The final angular mismatch (mod 2 $\pi$) is the total phase and corresponding colors relate to corresponding values from 0 to $2 \pi$.  Note that there are multiple curves where the phase is near 0.}
	\end{center}
\end{figure}

Thus curves of the same total phase foliate the Binet ellipsoid.  An example\footnote{We use $J s^2$ in the units of the moment of inertia as opposed to $kg \, m^2$ to make transparent the relationship between the energy ($J$), angular momentum ($J s$), moments of inertia ($J s^2$), and angular velocity ($s^{-1}$) in \Eqns{eqn:TK} and (\ref{eqn:GeometricPhase}).} of this can be seen in \Figref{fig:fig4} for the case of ($I_1$, $I_2$, $I_3$) = ($0.286827 \, J s^2$, $0.533256 \, J s^2$, $0.795844 \, J s^2$). Here the color of the curve corresponds to the total phase modulo $2 \pi$.  

%
%
%
%
%
%

In the next section, we check \Eqn{eqn:GeometricPhase} by evaluating \Eqn{eqn:GeometricPhase}, by calculating \Eqn{eqn:montgomery} numerically, and by solving for a rotation matrix after one period.

\subsection{Demonstration of formula}
For illustrative purposes, we choose a handle shape where $l_1 = 8$, $l_2 = 4$, and $w = 1$ as shown in \Figref{fig:fig1}.  We further choose the total mass (from a uniform density) to be such that the moment of inertia has unit magnitude.  The initial angular velocity is chosen so that it has unit magnitude, and the direction is given by a polar angle (measured from the body $z$ axis), $\theta_{\bd{\omega}} = \pi/2 -  0.15$, and an azimuthal angle, $\phi_{\bd{\omega}} = \pi/2 - 0.2$, measured from body $x$ axis.  This corresponds to a large initial velocity near the middle moment of inertia, $I_2$.  This sets the total angular momentum and total rotational energy.  Those, along with the moments of inertia are: $E = 0.264805 \, J$, $M = 0.533252 \, J s $, $I_1 = 0.286827 \, J s^2$, $I_2 = 0.533256 \, J s^2$, and $I_3 = 0.795844 \, J s^2$.  The dynamics of $\bd{\omega}$ in the body frame are shown in \Figref{fig:fig3}(b).  In this case, the period calculated from \Eqn{eqn:TK} (a) is 23.8181 s.  We can see from the shape of the curves that this corresponds visually to the period of \Figref{fig:fig3}(b).  One aspect to note from this plot is that the dashed green curve shows the body $y$-component of the angular velocity (the axis of the intermediate moment of inertia) suddenly reverses twice describing the flip known as the Dzhanibekov effect or the tennis racket theorem.\cite{ashbaugh1991twisting, levi2014classical, goldstein, LL}  Interestingly, most classic text in dynamics describe this behavior as `unstable', however the motion is completely determined and is periodic, eventually coming back to the `unstable' point.  

We may now compute some of the quantities that have been developed.  The dynamic part of the total phase from \Eqn{eqn:montgomery} $\left( \dfrac{2 E T}{M} \right)$ is 23.6555.  If we integrate the projected angular velocity (from \Eqn{eqn:Omegas} or numerically integrating \Eqn{eqn:Euler}) along the angular momentum axis for one period, we get
\begin{equation}
\int_{0}^{T} \frac{\bd{\omega} \cdot \bd{M}}{|\bd{M}|} \dd t = 23.6555,
\end{equation}
which matches the dynamic part.  If we compute the solid angle of intersection between the sphere of angular momentum and the Binet ellipsoid numerically using \emph{Mathematica} (seen in \Figref{fig:fig3} (a)), we get the signed area $\Omega = 2.1253$.

This gives a calculated total phase of 21.5302.  The formula from \Eqn{eqn:GeometricPhase} gives a value of 21.5303.  The difference is mostly due to the numerical discretization of the area of intersection and can be made to be smaller with a finer discretization.

Another way of checking both formulas is by numerically integrating the Euler equations over one period and solving a system of equations for the rotation matrix after one period.  The rotation matrix would give the geometric rotation of the initial system compared to the final orientation after one period.  In this case, the rotation around the angular momentum vector is given by 2.68074 which is precisely the value of the total phase formula (21.5303), modulo $2 \pi$.

To further simplify the total phase formula, we may use the observation that there are related initial conditions that describe the same total phase (since the total energy and magnitude of the angular momentum is the same throughout the evolution).  These lie on the same closed curve on the Binet ellipsoid (\Figref{fig:fig4}).  Without loss of generality, we may choose the initial conditions to lie on the space $x-z$ plane so that the azimuthal angle of the initial angular velocity, $\phi_{\bd{\omega}}$ is equal to 0.\footnote{It might be objected that by the time the initial condition ($\theta_{\bd{\omega}}$ and $\phi_{\bd{\omega}}$) makes it to the $x-z$ plane, the angular velocity may no longer be of unit magnitude (in fact, in general, it will not).  However since \Eqn{eqn:GeometricPhase} is independent of $|\bd{\omega}_0|$, we may renormalize and retain the same $\Delta \alpha$, but not necessarily the same exact dynamics.  Equivalently (as discussed in \S \ref{sec:sym}), we may rescale our unit of time to give $|\bd{\omega}_0| = 1$.  This requires scaling the time by the ratio of the periods.  The Binet ellipsoid and angular momentum sphere are simply magnified or reduced, but the polhode curve shape is maintained.}  Thus for the remainder of the paper we take 
\begin{equation}\label{eqn:omega0}
\bd{\omega}_0 = (\sin \theta_{\bd{\omega}},0,\cos \theta_{\bd{\omega}}).
\end{equation}

If we consider all combinations of moments of inertia in the region indicated in \Figref{fig:fig2}, along with all initial $0 < \theta_\omega < \pi/2$, this covers all initial conditions (using symmetry of the angle, direction of spin, and independence from $|\bd{\omega}_0|$ and $|\bd{I}|$).  Equation (\ref{eqn:GeometricPhase}) has been checked exhaustively against \Eqn{eqn:montgomery} for 292 sets of moments of inertia equally spaced throughout the region in \Figref{fig:fig2} with 22 angles between $0<\theta_{\bd{\omega}} < \pi/2$ for each moment of inertia choice.  This corresponds to 6446 initial conditions.  In all cases, both \Eqn{eqn:GeometricPhase} and \Eqn{eqn:montgomery} give the same value.

\section{Discussion of Formula}
\subsection{Symmetries of the total phase}\label{sec:sym}
As mentioned before, the total phase formula has notable symmetries including its independence from $|\bd{I}| = \sqrt{I_1^2+I_2^2+I_3^2}$ and $|\bd{\omega}_{0}| = \sqrt{\omega_{01}^2+\omega_{02}^2+\omega_{03}^2}$.  

Independence from $|\bd{I}|$ follows from the observation that $|\bd{I}|$ is directly proportional to the total mass of a rigid body.  Changing $|\bd{I}|$ amounts to changing the mass scale.  If the mass scale were changed, the dynamics would be unchanged, especially the geometric quality of the rotation.

To qualitatively understand the independence of \Eqn{eqn:GeometricPhase} from $|\bd{\omega}_{0}|$, we see that changing the magnitude of $|\bd{\omega}_{0}|$ changes the timescale for the dynamics, but geometrically, does not affect the phase.  If we had made a video recording of the spinning, and played it back at a slower speed, the geometrical quality of the rotation would also be unchanged.

Thus we only need three parameters to specify the total phase: two moments of inertia, $I_1$ and $I_2$ (or two ellipsoidal semi-axes ($b$ and $c$)), and the polar angle of the initial angular velocity in the body frame, $\theta_{\bd{\omega}}$.  

We may explicitly plug in the expressions for $E$ and $M$ (\Eqns{eqn:E} and (\ref{eqn:J})) along with \Eqn{eqn:omega0} and the definitions for $T$ and $k$ (\Eqns{eqn:TK}) into the total phase formula (\Eqn{eqn:GeometricPhase}) to give
\begin{widetext}
\begin{equation}\label{eqn:alphatheta}
\Delta \alpha =\begin{cases} 
      4 \sqrt{\beta \frac{I_2 \left( I_3^2 - I_1^2 (1- \sec^2 \theta_{\bd{\omega}}) \right)}{I_1^2 I_3 (I_3 - I_1) (I_2 - I_1)}} \left( I_1 \mathlarger{\textrm{K}}\! \left( \beta \tan^2 \theta_{\bd{\omega}} \right) + (I_3 - I_1) \mathlarger{\Pi}\!\left( -\frac{I_3^2}{I_1^2} \beta \big| \beta \tan^2 \theta_{\bd{\omega}} \right) \right) - 2 \pi & M^2 > 2 E I_2 \\
      4 \sqrt{\frac{I_2 \left( I_1^2 + I_3^2 \cot^2 \theta_{\bd{\omega}} \right)}{I_1^2 I_3 (I_3 - I_1) (I_2 - I_1)}} \left( I_1 \mathlarger{\textrm{K}}\! \left( \frac{\cot^2 \theta_{\bd{\omega}}}{\beta} \right) +(I_3 - I_1) \mathlarger{\Pi}\!\left( -\frac{I_3^2 \cot^2 \theta_{\bd{\omega}}}{I_1^2} \beta \big| \frac{\cot^2 \theta_{\bd{\omega}}}{\beta} \right) \right) & M^2 < 2 E I_2 
   \end{cases},
\end{equation}
\end{widetext}
where
\begin{equation}\label{eqn:beta}
\beta = \frac{I_1 (I_2 - I_1)}{I_3 (I_3 - I_1)},
\end{equation}
and $\mathlarger{\textrm{K}}$ and $\mathlarger{\Pi}$ are defined in \Eqns{eqn:K} and (\ref{eqn:Pi}), respectively. Here we might have eliminated $I_3$ as well using $I_3 = \sqrt{1-I_1^2 - I_2^2}$, but that would perhaps not benefit the exposition.

%
%
%
%
%
%
\begin{figure}[!hb]
	\begin{center}
 	\includegraphics[width=\figsize]{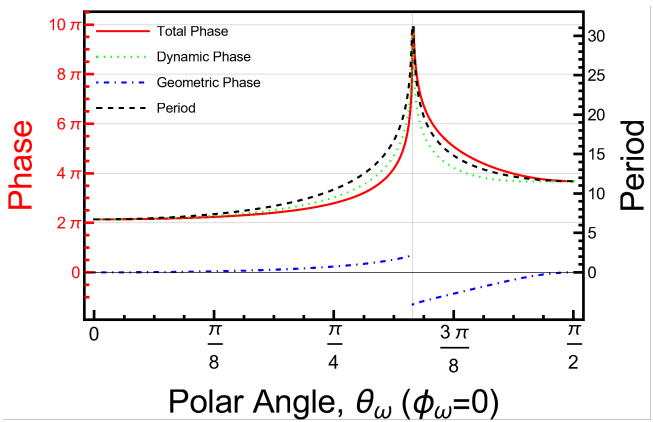}\\
      \caption{\label{fig:fig5}  For the demonstrated example of \Figref{fig:fig3}, the solid red curve indicates the total phase for different initial conditions with varying polar angle of the angular velocity.  A polar angle of 0 is an angular velocity along the body $z$ axis.  The total phase (solid red) is the difference between the dynamic phase (dotted green) and the geometric phase (dot-dashed blue).  Here we see that regardless of the initial condition, the total phase will be larger than about $2 \pi$ radians.  The black dashed curve plots the corresponding period of the angular velocity or angular momentum vector in seconds.}
	\end{center}
\end{figure}
\subsection{Minimum Total Phase}
Figure \ref{fig:fig5} shows two plots for the above demonstrated case.  The solid red curve is the total phase for a given initial condition described by the polar angle ($\theta_{\bd{\omega}}$) of the angular velocity with respect to the body $z$ axis and an azimuthal angle of zero ($\phi_{\bd{\omega}}$) with respect to the body $x$ axis. The total phase (solid red) is the difference between the dynamic phase (dotted green) and the geometric phase (dot-dashed blue).  The dashed black curve is the corresponding time period, $T$, for the angular velocity or angular momentum vectors.  For a given object, the total phase, dynamic phase, and period diverge at an angle that corresponds to the separatrix for the intersection of the Binet ellipsoid and the moment of inertia sphere.  The critical angle is given by the condition 
\begin{equation}
M^2 = 2 E I_2
\end{equation}
which reduces to 
\begin{equation}
\theta_{\textrm{crit}} = \cot^{-1} \left( \sqrt{ \beta } \right),
\end{equation}
with $\beta$ defined in \Eqn{eqn:beta}.

\begin{figure}[!hb]
	\begin{center}
 	\includegraphics[width=\figsize]{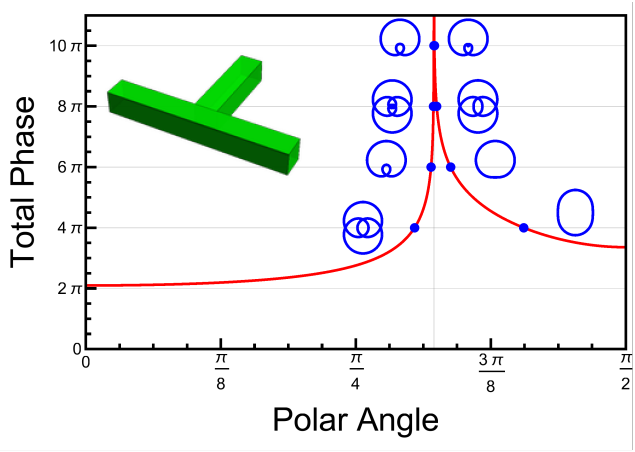}\\
      \caption{\label{fig:fig6} Total phase for moments of inertia given by ($I_1$, $I_2$, $I_3$) = (0.26 $\, J s^2$, 0.55$\, J s^2$, 0.793662$\, J s^2$).  Initial angular velocity angle is given as a polar angle from the body $z$ axis.  When the total phase is an integer multiple of 2 $\pi$, the herpolhode (curve of the terminus of the angular velocity in space coordinates) is a close plane curve.  Various examples are plotted near the intersection of multiples of $2 \pi$ and the total phase curve.}
	\end{center}
\end{figure}

This critical angle is displayed as a gray vertical line on \Figrefs{fig:fig5} and \ref{fig:fig6}.  Note that the minimum value for the total phase occurs at the extremes of 0 and $\pi$.  This appears to be a generic behavior in all of the initial conditions investigated.  It might not be too difficult to differentiate \Eqns{eqn:alphatheta} and investigate the sign of the derivative, but may not add much.  Additionally, near the critical angle, the total phase diverges.

When the total phase modulo $2 \pi$ is equal to 0, trajectories that are started on these curves are closed and each Euler angle (mod $2 \pi$) is periodic with a period whose integer multiple is the period of the angular velocity vector.  As the trajectories on the Binet ellipsoid approach the curve $M^2 = 2 E I_2$, the total phase increases without bound.  One question that is apparent is what is the minimum total phase for a given object?  For example are there objects ($I_1$ and $I_2$ or $b$ and $c$ values) where there are no initial conditions that give rise to a total phase of exactly $2 \pi$?

In \Figref{fig:fig2} we see that the minimum total phase can be as high as desired.  Looking at all objects, the largest minimum total phase is unbounded for objects with moments of inertia approaching $I_1 = I_2 = I_3 = \sqrt{3}/3$.  There is an apparent paradox in this behavior.  If all moments of inertia are equal and the total angular momentum is fixed, then the body spins around whatever axis it is initial spinning about.  The total phase would seem to be zero.  However the total phase formula goes to infinity.  The resolution may be seen in the behavior of the period $T$.  When the moments of inertia each approach $\sqrt{3}/3$, the period is seen to diverge.  Thus with $T$ going to infinity, the total phase is permitted to correspondingly diverge.  Note that when $I_1 = I_2 = I_3 = \sqrt{3}/3$ exactly, the object is no longer dynamically asymmetric (an example of this is a sphere). 

On the other hand, the asymmetric top with the smallest minimum possible total phase for any initial condition is a total phase of 0 for objects with moments of inertia approaching $I_1 = 0$ and $I_2 = I_3 = \sqrt{2}/2$.  An example of this is a thin solid rod.

If we look at initial conditions near the body $z$ or $x$ axes ($\theta_{\bd{\omega}} \approx 0$ or $\theta_{\bd{\omega}} \approx \pi/2$, respectively), we may work out the minimum total phase directly from linearizing the Euler equations (\Eqns{eqn:Euler}). The Euler equations reduce to a simple harmonic oscillator equation where the period may be identified directly.  Since in this case, the geometric part of the total phase is zero (the angular momentum does not enclose any solid angle), the total phase is given by the dynamic part, for which there is a simple formula (see \Eqn{eqn:montgomery}).

In the case where $\theta_{\bd{\omega}} \approx 0$ the minimum total phase reduces to
\begin{equation}\label{eqn:alphamin1}
\Delta \alpha_{\textrm{min,z}} = 2 \pi \sqrt{\frac{I_1 I_2}{(I_3 - I_2) (I_3 - I_1)}}.
\end{equation}
In the case where $\theta_{\bd{\omega}} \approx \pi/2$ the minimum total phase is given by
\begin{equation}\label{eqn:alphamin2}
\Delta \alpha_{\textrm{min,z}} = 2 \pi \sqrt{\frac{I_3 I_2}{(I_2 - I_1) (I_3 - I_1)}}.
\end{equation}
These formulas may also be derived by expanding \Eqn{eqn:alphatheta} near $\theta_{\bd{\omega}} \approx 0$ and $\theta_{\bd{\omega}} \approx \pi/2$.  It is worth noting that the total phase near these extremes is composed of entirely the dynamic part of the total phase.  For initial conditions away from these extremes, the total phase is a combination of the dynamic and geometric parts as can be seen in \Figref{fig:fig5}.

The locus of points where \Eqn{eqn:alphamin1} and \Eqn{eqn:alphamin2} are equal is exactly the location of the kinks in the contours of \Figref{fig:fig2}.
%
%
%
%
%
%
%
\subsection{Closed Herpolhodes}
We may also use the total phase formula (\Eqn{eqn:GeometricPhase}) to produce curves of closed herpolhodes.  A herpolhode is the curve traced out by the terminus of the angular velocity vector in the space coordinates.  In general these do not close after a period.  However if we use initial conditions that have a total phase that is an integer multiple of $2 \pi$, we find interesting closed curve patterns.  

Figure \ref{fig:fig6} shows the total phase for all initial polar angles ($\theta_{\bd{\omega}}$) of the angular velocity (with $\phi_{\bd{\omega}} = 0$). The particular object (whose moments of inertia are given by ($I_1$, $I_2$, $I_3$) = (0.26$\, J s^2$, 0.55$\, J s^2$, 0.793662$\, J s^2$)) has initial conditions with closed herpolhode trajectories.  The corresponding angles that result in the smallest closed herpolhodes are (from left to right) 0.9562, 1.004, 1.011, 1.012, 1.014, 1.021, 1.061, 1.2742 radians. These are exactly the angles where the total phase is a multiple of 2$\pi$.  In this case the multiples are 2, 3, 4, 5, 5, 4, 3, and 2 respectively.  Note that closed herpolhodes are possible for all integer multiples of $2 \pi$ greater than 2.  The critical angle is 1.01262.  Some of the closed herpolhode figures are depicted in \Figref{fig:fig6}.

Two other comments about closed herpolhodes should be made.  There is vertical symmetry for all herpolhodes in the plot, but for the even multiples of 2 $\pi$, ($4\pi$, $8\pi$, $12\pi$, etc.), the herpolhodes also have a horizontal symmetry.  Additionally, changing the magnitude of $|\bd{\omega}_0|$ expands or contracts the herpolhode image, but otherwise keeps it unchanged.

It is unclear what, if any, connection these closed curves and the associated energies have to the Bohr-Sommerfeld quantization in quantum mechanics.

\section{Applications and Possible Connections}
The primary purpose of this paper was to use the asymmetric top as an intuitive and natural example of total phase.  However we give three examples of the connections that could be made with other areas of physics.

Quantization of the asymmetric top\cite{postell1990quantization} leads to quantized energy levels of the quantum mechanical asymmetric top.  How this corresponds to the closed orbits of the last section still needs to be studied.

When molecules are no longer symmetric, the asymmetry leads to observable spectroscopic modes that allow for analysis and characterization.\cite{blaker, herzberg1945infrared, townes2013microwave} (See Ch.\ 1, \S 4 of Ref.\ (\onlinecite{herzberg1945infrared}) and Ch.\ 4 of Ref.\ (\onlinecite{townes2013microwave}))  Understanding the dynamics of the asymmetric top classically gives us an understanding of the connection with the rotational modes of an asymmetric molecule.  Making further connections with the total phase arising from simple rotation of these molecules could lead to further insight.

Large nuclei such as $\phantom{}^{135} \textrm{Pr}$ and $\phantom{}^{163} \textrm{Lu}$, provide examples of asymmetric top systems in nuclear physics.\cite{PhysRevC.89.014322, odegaard2001evidence}  Through X-ray spectroscopy, the so-called `wobble' bands provide a way to measure angular momentum transitions of spinning nuclei.  The connection to closed total phase orbits still needs to be made.

\section{Conclusions}
In this paper, we have found the total phase formula (and thus trivially the geometric phase) explicitly in terms of special functions.  This leads immediately to observations about the total phase for the asymmetric top in force-free motion.  The total phase of a rigid object is independent of the magnitude of the moment of inertia, independent of the magnitude of the angular velocity, and all initial conditions along the path of the angular velocity vector in space have the same total phase and geometric phase.  Thus for a given rigid body, the total phase, dynamic phase, and geometric phase only depend on a single angle, the polar angle that the initial angular velocity vector makes when it crosses the $x-z$ plane (for instance).  Additionally, in the parameter space of the moments of inertia, each class of asymmetric rigid bodies has a minimum total phase; that is, a minimum angle by which an object must rotate after undergoing one period of the angular velocity vector in space.  For some objects, the total phase can be as high as desired.  Closed trajectories were also highlighted and closed herpolhodes were illustrated and discussed briefly.

\begin{acknowledgments}
The author would like to thank Richard Cecil for many conversations relating to geometric phase and other topics.  Additionally, the author would like to thank the input of Richard Montgomery for many observations and suggesting the line of analysis for \Eqns{eqn:alphamin1} and (\ref{eqn:alphamin2}).  This work was funded by the Vitreous State Laboratory.
\end{acknowledgments}

\appendix*
\section{Explicit Derivation of the Time Dependence of the Euler Angles}
The solution of the time dependence of the angular velocity for an asymmetric top in force-free motion is given in Eq.\ 37.10 of Ref.\ \onlinecite{LL}, reproduced here for convenience:
\begin{subequations}\label{eqn:Omegas}
\begin{align}
\omega_1(t) &= \sqrt{\frac{2 E I_3 - M^2}{I_1 (I_3 - I_1)}} \, \textrm{cn} (D t | k^2) \\
\omega_2(t) &= \sqrt{\frac{2 E I_3 - M^2}{I_2 (I_3 - I_2)}} \, \textrm{sn} (D t | k^2) \\
\omega_3(t) &= \sqrt{\frac{M^2 - 2 E I_1}{I_3 (I_3 - I_1)}} \, \textrm{dn} (D t | k^2).
\end{align}
\end{subequations}
Here $\textrm{sn} (u | k^2)$ is the Jacobi Elliptic Sine function defined by inverting the Jacobi Elliptic Integral of the First Kind, $\mathlarger{\textrm{F}}$,  through the pair of equations
\begin{subequations}\label{eqn:sn}
\begin{align}
\sin \phi &= \textrm{sn} (u|k^2) \\
u &= \mathlarger{\textrm{F}}\!(\phi | k^2) =  \int_{0}^{\phi} \frac{\dd t}{\sqrt{1- k^2 \sin^2 t}},
\end{align}
\end{subequations}
and $\textrm{cn} (u | k^2) = \sqrt{1-\textrm{sn}^2 (u | k^2)}$ and $\textrm{dn} (u | k^2) = \sqrt{1-k^2 \textrm{sn}^2 (u | k^2)}$.  The definition of $D$ is given in \Eqn{eqn:ABCD}(d).  

The above equations (\Eqns{eqn:Omegas}) are valid when $M^2>2 E I_2$.  When $M^2<2 E I_2$, all indices $1 \leftrightarrow 3$ (including in the definition of $k$ and $\omega$).  Alternatively, we may use \Eqns{eqn:Omegas} for \emph{all} initial conditions without making any index changes (extending $k>1$ in the region $M^2<2 E I_2$).  This is the most convenient, but care must be taken.  For the rest of the Appendix, we use this extended definition of $k$ to simplify the notation but at the end of the Appendix produce equations that are valid in the notation of the text.  To distinguish it, we use $\kappa$.  Thus
\begin{equation}\label{eqn:kappa}
\kappa = \sqrt{\frac{(I_2 - I_1) (2 E I_3 - M^2)}{(I_3 - I_2) (M^2 - 2 E I_1)}},
\end{equation}
for all $I_1<I_2<I_3$ and any $E$ and $M$.

We may further express the Euler angles as explicit functions of time.  We choose the Euler angle convention of both Refs.\ (\onlinecite{goldstein}) and (\onlinecite{LL}) (Figure 4.7, p.\ 152 and Fig.\ 47 \S 35, respectively).  This describes a sequence of intrinsic elemental rotations about the body $z$-$x'$-$z''$ axes. In this convention, two of the Euler angles ($\theta$ and $\psi$) are given in terms of the angular velocity functions (see Ref.\ \onlinecite{LL}, Eq.\ 37.14), 
\begin{subequations}\label{eqn:eulerangles}
\begin{align}
\cos \theta &= \frac{I_3 \omega_3 }{M}   \\
\tan \psi  &= \frac{I_1 \omega_1 }{I_2 \omega_2}.
\end{align}
\end{subequations}
%
%
%
%
%
%

The final Euler angle, $\phi$, is given as a first-order non-linear differential equation as follows.

Using Eq.\ 37.16 of Ref.\ \onlinecite{LL}, we may substitute in the expressions of Eq.\ 37.6 of Ref.\ \onlinecite{LL}.  Simplifying and using the explicit expressions of the angular velocity in terms of Jacobi elliptic functions (Eq.\ 37.10, Ref.\ \onlinecite{LL} and above \Eqns{eqn:Omegas}), we arrive at the expression of the time rate of change of $\phi$, 
\begin{equation}
\frac{\dd \phi}{\dd t} = \frac{M}{I_3} + \frac{A}{B + C \, \textrm{sn}^2 \left( D t | \kappa^2 \right)}
\end{equation}
with 
\begin{subequations}\label{eqn:ABCD}
\begin{align}
A &= - M (I_1 + I_3)(I_3 - I_2) < 0 \\
B &= I_1 I_3 (I_3 - I_2) > 0 \\
C &= I_3^2 (I_2 - I_1) > 0 \\
D &= \sqrt{\frac{(I_3 - I_2) (M^2 - 2 E I_1)}{I_1 I_2 I_3}},
\end{align}
\end{subequations}
and $\kappa$ defined in \Eq{eqn:kappa}.  This is valid for all initial conditions.  

When we integrate this, we get 
\begin{equation}
\phi = \int_{0}^{t} \frac{\dd \phi}{\dd t} \dd t = \frac{M}{I_3} t + \left( \frac{A}{B D} \right) \int_{0}^{D t} \frac{\dd u }{1 - \alpha^2 \, \textrm{sn}^2 \left( u | \kappa^2 \right)},
\end{equation}
with 
\begin{equation}
\alpha^2 = -\frac{C}{B}.
\end{equation}
We may use Eq.\ 400.01 (in the definition of the Incomplete Elliptic Integral of the Third Kind of Ref.\ \onlinecite{byrdhandbook}) 
\begin{equation}
\mathlarger{\Pi}\!\!\left( \alpha^2 ; \phi \big| \kappa^2 \right) = \int_{0}^{u} \frac{\dd x }{1 - \alpha^2 \, \textrm{sn}^2 \left( x | \kappa^2 \right)},
\end{equation}
where $u$ is defined in \Eq{eqn:sn} (a), to do the above integral.  The exact expressions for each region of initial conditions ($M^2 > 2 E I_2$ and $M^2 < 2 E I_2$) must be done with care to avoid problems with the definitions of built-in functions.  

For the case where initial conditions give $M^2 > 2 E I_2$ ($0<\kappa<1$), we get 
\begin{align}\label{eqn:phi1}
&\phi (t) = \frac{M}{I_3} t \nonumber\\
&+ \frac{M (I_3 - I_1)}{I_1 I_3 D} \mathlarger{\Pi}\!\!\left( - \frac{I_3 (I_2 - I_1)}{I_1 (I_3 - I_2)}; \mathlarger{\text{am}}\! (D t | \kappa^2) \big| \kappa^2 \right),
\end{align}
where $\mathlarger{\text{am}}\!\!\left( u \big| m \right)$ is the Jacobi Amplitude.  If $u = \mathlarger{\textrm{F}}\!\left(\phi \big| m \right)$ then $\phi =  \mathlarger{\text{am}}\!\!\left( u \big| m \right)$, where $\mathlarger{\textrm{F}}$ is the Incomplete Elliptic Integral of the First Kind defined in \Eqn{eqn:sn} (b). Again, $\kappa$ is defined in \Eq{eqn:kappa}, and $D$ is given in \Eqn{eqn:ABCD}(d).

For the case $M^2 < 2 E I_2$ ($\kappa > 1$) we must use the reciprocal modulus transformation on the Incomplete Elliptic Integral of the Third Kind (19.7.4 of Ref.\ \onlinecite{olver2010nist}, p.\ 492 or 162.02 p.\ 39 of Ref.\ \onlinecite{byrdhandbook}) as well as the reciprocal modulus transformation for the Jacobi Elliptic Sine function (162.01 p.\ 39 of Ref.\ \onlinecite{byrdhandbook}) and definition 22.16.1 of Ref.\ \onlinecite{olver2010nist} to give
\begin{align}\label{eqn:phi2}
&\phi (t) = \frac{M}{I_3} t \nonumber\\
&+ \frac{M (I_3 - I_1)}{I_1 I_3 D} \frac{1}{\kappa} \mathlarger{\Pi}\!\!\left( - \frac{I_3 (I_2 - I_1)}{\kappa^2 I_1 (I_3 - I_2)}; \mathlarger{\text{am}}\! (\kappa D t | \kappa^{-2}) \big| \kappa^{-2} \right).
\end{align}
As before $\kappa$ is defined in \Eq{eqn:kappa}, and $D$ is given in \Eqn{eqn:ABCD}(d).

Note that in the region $M^2 < 2 E I_2$, $\kappa = 1/k$ with k defined through \Eqn{eqn:TK} (b).  Thus we may rewrite \Eqns{eqn:phi1} and (\ref{eqn:phi2}) with notation in the rest of the paper as
%
%
%
%
%
%
%
%
\begin{widetext}
\begin{equation}\label{eqn:fullphi}
\phi(t) = \begin{cases} 
      \frac{M}{I_3} t + \frac{M (I_3 - I_1)}{I_1 I_3 D} \mathlarger{\Pi}\!\left( - \frac{I_3 (I_2 - I_1)}{I_1 (I_3 - I_2)}; \mathlarger{\text{am}}\! (D t | k^2) \big| k^2 \right) & M^2 > 2 E I_2 \\
      \frac{M}{I_3} t + \frac{M (I_3 - I_1)}{I_1 I_3 D} k \mathlarger{\Pi}\!\left( - \frac{k^2 I_3 (I_2 - I_1)}{I_1 (I_3 - I_2)}; \mathlarger{\text{am}}\! \left(\frac{D t}{k} | k^2 \right) \big| k^2 \right) & M^2 < 2 E I_2 
   \end{cases}.
\end{equation}
\end{widetext}


\begin{thebibliography}{35}
\expandafter\ifx\csname natexlab\endcsname\relax\def\natexlab#1{#1}\fi
\expandafter\ifx\csname bibnamefont\endcsname\relax
  \def\bibnamefont#1{#1}\fi
\expandafter\ifx\csname bibfnamefont\endcsname\relax
  \def\bibfnamefont#1{#1}\fi
\expandafter\ifx\csname citenamefont\endcsname\relax
  \def\citenamefont#1{#1}\fi
\expandafter\ifx\csname url\endcsname\relax
  \def\url#1{\texttt{#1}}\fi
\expandafter\ifx\csname urlprefix\endcsname\relax\def\urlprefix{URL }\fi
\providecommand{\bibinfo}[2]{#2}
\providecommand{\eprint}[2][]{\url{#2}}

\bibitem[{\citenamefont{{Plasma Ben}}()}]{thandle}
\bibinfo{author}{\bibnamefont{{Plasma Ben}}}, \emph{\bibinfo{title}{{D}ancing
  {T}-handle in zero-g, {HD}}},
  \urlprefix\url{https://www.youtube.com/watch?v=1n-HMSCDYtM}.

\bibitem[{\citenamefont{Moffatt}(2000)}]{moffatt2000euler}
\bibinfo{author}{\bibfnamefont{H.}~\bibnamefont{Moffatt}},
  \bibinfo{journal}{Nature} \textbf{\bibinfo{volume}{404}},
  \bibinfo{pages}{833} (\bibinfo{year}{2000}).

\bibitem[{\citenamefont{Jalali et~al.}(2015)\citenamefont{Jalali, Sarebangholi,
  and Alam}}]{rollingrings}
\bibinfo{author}{\bibfnamefont{M.~A.} \bibnamefont{Jalali}},
  \bibinfo{author}{\bibfnamefont{M.~S.} \bibnamefont{Sarebangholi}},
  \bibnamefont{and} \bibinfo{author}{\bibfnamefont{M.-R.} \bibnamefont{Alam}},
  \bibinfo{journal}{Physical Review E} \textbf{\bibinfo{volume}{92}},
  \bibinfo{pages}{032913} (\bibinfo{year}{2015}).

\bibitem[{\citenamefont{Berry and Shukla}(2010)}]{berry2010slow}
\bibinfo{author}{\bibfnamefont{M.}~\bibnamefont{Berry}} \bibnamefont{and}
  \bibinfo{author}{\bibfnamefont{P.}~\bibnamefont{Shukla}},
  \bibinfo{journal}{European Journal of Physics} \textbf{\bibinfo{volume}{32}},
  \bibinfo{pages}{115} (\bibinfo{year}{2010}).

\bibitem[{\citenamefont{Berry and Hannay}(1988)}]{berry1988classical}
\bibinfo{author}{\bibfnamefont{M.}~\bibnamefont{Berry}} \bibnamefont{and}
  \bibinfo{author}{\bibfnamefont{J.}~\bibnamefont{Hannay}},
  \bibinfo{journal}{Journal of Physics A: Mathematical and General}
  \textbf{\bibinfo{volume}{21}}, \bibinfo{pages}{L325} (\bibinfo{year}{1988}).

\bibitem[{\citenamefont{Montgomery}(1991)}]{montgomery1991much}
\bibinfo{author}{\bibfnamefont{R.}~\bibnamefont{Montgomery}},
  \bibinfo{journal}{Am. J. Phys} \textbf{\bibinfo{volume}{59}},
  \bibinfo{pages}{394} (\bibinfo{year}{1991}).

\bibitem[{\citenamefont{Hannay}(1985)}]{hannay1985angle}
\bibinfo{author}{\bibfnamefont{J.~H.} \bibnamefont{Hannay}},
  \bibinfo{journal}{Journal of Physics A: Mathematical and General}
  \textbf{\bibinfo{volume}{18}}, \bibinfo{pages}{221} (\bibinfo{year}{1985}).

\bibitem[{\citenamefont{Xiao et~al.}(2010)\citenamefont{Xiao, Chang, and
  Niu}}]{xiao2010berry}
\bibinfo{author}{\bibfnamefont{D.}~\bibnamefont{Xiao}},
  \bibinfo{author}{\bibfnamefont{M.-C.} \bibnamefont{Chang}}, \bibnamefont{and}
  \bibinfo{author}{\bibfnamefont{Q.}~\bibnamefont{Niu}},
  \bibinfo{journal}{Reviews of modern physics} \textbf{\bibinfo{volume}{82}},
  \bibinfo{pages}{1959} (\bibinfo{year}{2010}).

\bibitem[{\citenamefont{Resta}(2000)}]{resta2000manifestations}
\bibinfo{author}{\bibfnamefont{R.}~\bibnamefont{Resta}},
  \bibinfo{journal}{Journal of Physics: Condensed Matter}
  \textbf{\bibinfo{volume}{12}}, \bibinfo{pages}{R107} (\bibinfo{year}{2000}).

\bibitem[{\citenamefont{Mead}(1992)}]{mead1992geometric}
\bibinfo{author}{\bibfnamefont{C.~A.} \bibnamefont{Mead}},
  \bibinfo{journal}{Reviews of modern physics} \textbf{\bibinfo{volume}{64}},
  \bibinfo{pages}{51} (\bibinfo{year}{1992}).

\bibitem[{\citenamefont{Berry}(1984)}]{berry1984}
\bibinfo{author}{\bibfnamefont{M.~V.} \bibnamefont{Berry}},
  \bibinfo{journal}{Proceedings of the Royal Society of London. A. Mathematical
  and Physical Sciences} \textbf{\bibinfo{volume}{392}}, \bibinfo{pages}{45}
  (\bibinfo{year}{1984}).

\bibitem[{\citenamefont{Anandan}(1992)}]{anandan1992geometric}
\bibinfo{author}{\bibfnamefont{J.}~\bibnamefont{Anandan}},
  \bibinfo{journal}{Nature} \textbf{\bibinfo{volume}{360}},
  \bibinfo{pages}{307} (\bibinfo{year}{1992}).

\bibitem[{\citenamefont{Berry}(1988)}]{berry1988geometric}
\bibinfo{author}{\bibfnamefont{M.}~\bibnamefont{Berry}},
  \bibinfo{journal}{Scientific American} \textbf{\bibinfo{volume}{259}},
  \bibinfo{pages}{46} (\bibinfo{year}{1988}).

\bibitem[{\citenamefont{Landau and Lifshitz}(1976)}]{LL}
\bibinfo{author}{\bibfnamefont{L.}~\bibnamefont{Landau}} \bibnamefont{and}
  \bibinfo{author}{\bibfnamefont{E.}~\bibnamefont{Lifshitz}},
  \emph{\bibinfo{title}{Mechanics}}, vol.~\bibinfo{volume}{1} of
  \emph{\bibinfo{series}{Course of Theoretical Physics}}
  (\bibinfo{year}{1976}).

\bibitem[{\citenamefont{Lawson and Rave}(2016)}]{lawson2016spacewalks}
\bibinfo{author}{\bibfnamefont{J.}~\bibnamefont{Lawson}} \bibnamefont{and}
  \bibinfo{author}{\bibfnamefont{M.}~\bibnamefont{Rave}},
  \bibinfo{journal}{Mathematics Magazine} \textbf{\bibinfo{volume}{89}},
  \bibinfo{pages}{105} (\bibinfo{year}{2016}).

\bibitem[{\citenamefont{Gil}(2010)}]{gil2010mechanical}
\bibinfo{author}{\bibfnamefont{S.}~\bibnamefont{Gil}},
  \bibinfo{journal}{American Journal of Physics} \textbf{\bibinfo{volume}{78}},
  \bibinfo{pages}{384} (\bibinfo{year}{2010}).

\bibitem[{\citenamefont{Jos{\'e} and Saletan}(2000)}]{jose2000classical}
\bibinfo{author}{\bibfnamefont{J.}~\bibnamefont{Jos{\'e}}} \bibnamefont{and}
  \bibinfo{author}{\bibfnamefont{E.}~\bibnamefont{Saletan}},
  \emph{\bibinfo{title}{Classical dynamics: a contemporary approach}}
  (\bibinfo{year}{2000}).

\bibitem[{\citenamefont{Marsden et~al.}(1992)}]{marsden1992lectures}
\bibinfo{author}{\bibfnamefont{J.~E.} \bibnamefont{Marsden}}
  \bibnamefont{et~al.}, \emph{\bibinfo{title}{Lectures on mechanics}}, vol.
  \bibinfo{volume}{174} (\bibinfo{publisher}{Cambridge University Press},
  \bibinfo{year}{1992}).

\bibitem[{\citenamefont{Robbins}(2016)}]{robbins2016hannay}
\bibinfo{author}{\bibfnamefont{J.}~\bibnamefont{Robbins}},
  \bibinfo{journal}{Journal of Physics A: Mathematical and Theoretical}
  \textbf{\bibinfo{volume}{49}}, \bibinfo{pages}{431002}
  (\bibinfo{year}{2016}).

\bibitem[{\citenamefont{Hart et~al.}(1987)\citenamefont{Hart, Miller, and
  Mills}}]{hart1987simple}
\bibinfo{author}{\bibfnamefont{J.~B.} \bibnamefont{Hart}},
  \bibinfo{author}{\bibfnamefont{R.~E.} \bibnamefont{Miller}},
  \bibnamefont{and} \bibinfo{author}{\bibfnamefont{R.~L.} \bibnamefont{Mills}},
  \bibinfo{journal}{American Journal of Physics} \textbf{\bibinfo{volume}{55}},
  \bibinfo{pages}{67} (\bibinfo{year}{1987}).

\bibitem[{\citenamefont{Levi}(1993)}]{levi1993geometric}
\bibinfo{author}{\bibfnamefont{M.}~\bibnamefont{Levi}},
  \bibinfo{journal}{Archive for rational mechanics and analysis}
  \textbf{\bibinfo{volume}{122}}, \bibinfo{pages}{213} (\bibinfo{year}{1993}).

\bibitem[{\citenamefont{Bates et~al.}(2005)\citenamefont{Bates, Cushman, and
  Savev}}]{bates2005rotation}
\bibinfo{author}{\bibfnamefont{L.}~\bibnamefont{Bates}},
  \bibinfo{author}{\bibfnamefont{R.}~\bibnamefont{Cushman}}, \bibnamefont{and}
  \bibinfo{author}{\bibfnamefont{E.}~\bibnamefont{Savev}},
  \bibinfo{journal}{Zeitschrift f{\"u}r Angewandte Mathematik und Physik
  (ZAMP)} \textbf{\bibinfo{volume}{56}}, \bibinfo{pages}{183}
  (\bibinfo{year}{2005}).

\bibitem[{\citenamefont{Goldstein et~al.}(2002)\citenamefont{Goldstein, Poole,
  and Safko}}]{goldstein}
\bibinfo{author}{\bibfnamefont{H.}~\bibnamefont{Goldstein}},
  \bibinfo{author}{\bibfnamefont{C.}~\bibnamefont{Poole}}, \bibnamefont{and}
  \bibinfo{author}{\bibfnamefont{J.}~\bibnamefont{Safko}},
  \emph{\bibinfo{title}{Classical Mechanics}} (\bibinfo{publisher}{Addison
  Wesley, San Francisco}, \bibinfo{year}{2002}).

\bibitem[{\citenamefont{Whittaker}(1988)}]{whittaker1988treatise}
\bibinfo{author}{\bibfnamefont{E.~T.} \bibnamefont{Whittaker}},
  \emph{\bibinfo{title}{A treatise on the analytical dynamics of particles and
  rigid bodies}} (\bibinfo{publisher}{Cambridge University Press},
  \bibinfo{year}{1988}).

\bibitem[{\citenamefont{Zwanziger et~al.}(1990)\citenamefont{Zwanziger, Koenig,
  and Pines}}]{zwanziger1990berry}
\bibinfo{author}{\bibfnamefont{J.~W.} \bibnamefont{Zwanziger}},
  \bibinfo{author}{\bibfnamefont{M.}~\bibnamefont{Koenig}}, \bibnamefont{and}
  \bibinfo{author}{\bibfnamefont{A.}~\bibnamefont{Pines}},
  \bibinfo{journal}{Annual Review of Physical Chemistry}
  \textbf{\bibinfo{volume}{41}}, \bibinfo{pages}{601} (\bibinfo{year}{1990}).

\bibitem[{\citenamefont{Ashbaugh et~al.}(1991)\citenamefont{Ashbaugh, Chicone,
  and Cushman}}]{ashbaugh1991twisting}
\bibinfo{author}{\bibfnamefont{M.~S.} \bibnamefont{Ashbaugh}},
  \bibinfo{author}{\bibfnamefont{C.~C.} \bibnamefont{Chicone}},
  \bibnamefont{and} \bibinfo{author}{\bibfnamefont{R.~H.}
  \bibnamefont{Cushman}}, \bibinfo{journal}{Journal of Dynamics and
  differential Equations} \textbf{\bibinfo{volume}{3}}, \bibinfo{pages}{67}
  (\bibinfo{year}{1991}).

\bibitem[{\citenamefont{Levi}(2014)}]{levi2014classical}
\bibinfo{author}{\bibfnamefont{M.}~\bibnamefont{Levi}},
  \emph{\bibinfo{title}{Classical mechanics with calculus of variations and
  optimal control: an intuitive introduction}}, vol.~\bibinfo{volume}{69}
  (\bibinfo{publisher}{American Mathematical Soc.}, \bibinfo{year}{2014}).

\bibitem[{\citenamefont{Postell and Uzer}(1990)}]{postell1990quantization}
\bibinfo{author}{\bibfnamefont{V.}~\bibnamefont{Postell}} \bibnamefont{and}
  \bibinfo{author}{\bibfnamefont{T.}~\bibnamefont{Uzer}},
  \bibinfo{journal}{Physical Review A} \textbf{\bibinfo{volume}{41}},
  \bibinfo{pages}{4035} (\bibinfo{year}{1990}).

\bibitem[{\citenamefont{Blaker et~al.}(1962)\citenamefont{Blaker, Sidran, and
  Kaercher}}]{blaker}
\bibinfo{author}{\bibfnamefont{J.~W.} \bibnamefont{Blaker}},
  \bibinfo{author}{\bibfnamefont{M.}~\bibnamefont{Sidran}}, \bibnamefont{and}
  \bibinfo{author}{\bibfnamefont{A.}~\bibnamefont{Kaercher}},
  \bibinfo{type}{Tech. Rep.}, \bibinfo{institution}{Grumman Aircraft
  Engineering Corp Bethpage NY, Research Dept} (\bibinfo{year}{1962}).

\bibitem[{\citenamefont{Herzberg}(1945)}]{herzberg1945infrared}
\bibinfo{author}{\bibfnamefont{G.}~\bibnamefont{Herzberg}},
  \emph{\bibinfo{title}{Infrared and Raman spectra of polyatomic molecules}}
  (\bibinfo{publisher}{D. Van Nostrand Company.; New York},
  \bibinfo{year}{1945}).

\bibitem[{\citenamefont{Townes and Schawlow}(2013)}]{townes2013microwave}
\bibinfo{author}{\bibfnamefont{C.~H.} \bibnamefont{Townes}} \bibnamefont{and}
  \bibinfo{author}{\bibfnamefont{A.~L.} \bibnamefont{Schawlow}},
  \emph{\bibinfo{title}{Microwave spectroscopy}} (\bibinfo{publisher}{Courier
  Corporation}, \bibinfo{year}{2013}).

\bibitem[{\citenamefont{Frauendorf and D\"onau}(2014)}]{PhysRevC.89.014322}
\bibinfo{author}{\bibfnamefont{S.}~\bibnamefont{Frauendorf}} \bibnamefont{and}
  \bibinfo{author}{\bibfnamefont{F.}~\bibnamefont{D\"onau}},
  \bibinfo{journal}{Phys. Rev. C} \textbf{\bibinfo{volume}{89}},
  \bibinfo{pages}{014322} (\bibinfo{year}{2014}).

\bibitem[{\citenamefont{{\O}deg{\aa}rd
  et~al.}(2001)\citenamefont{{\O}deg{\aa}rd, Hagemann, Jensen, Bergstroem,
  Herskind, Sletten, Toermaenen, Wilson, Tj{\o}m, Hamamoto
  et~al.}}]{odegaard2001evidence}
\bibinfo{author}{\bibfnamefont{S.}~\bibnamefont{{\O}deg{\aa}rd}},
  \bibinfo{author}{\bibfnamefont{G.~B.} \bibnamefont{Hagemann}},
  \bibinfo{author}{\bibfnamefont{D.~R.} \bibnamefont{Jensen}},
  \bibinfo{author}{\bibfnamefont{M.}~\bibnamefont{Bergstroem}},
  \bibinfo{author}{\bibfnamefont{B.}~\bibnamefont{Herskind}},
  \bibinfo{author}{\bibfnamefont{G.}~\bibnamefont{Sletten}},
  \bibinfo{author}{\bibfnamefont{S.}~\bibnamefont{Toermaenen}},
  \bibinfo{author}{\bibfnamefont{J.}~\bibnamefont{Wilson}},
  \bibinfo{author}{\bibfnamefont{P.}~\bibnamefont{Tj{\o}m}},
  \bibinfo{author}{\bibfnamefont{I.}~\bibnamefont{Hamamoto}},
  \bibnamefont{et~al.}, \bibinfo{journal}{Physical review letters}
  \textbf{\bibinfo{volume}{86}}, \bibinfo{pages}{5866} (\bibinfo{year}{2001}).

\bibitem[{\citenamefont{Byrd and Friedman}(1971)}]{byrdhandbook}
\bibinfo{author}{\bibfnamefont{P.}~\bibnamefont{Byrd}} \bibnamefont{and}
  \bibinfo{author}{\bibfnamefont{M.}~\bibnamefont{Friedman}},
  \bibinfo{journal}{Berlin, Heidelberg. New York}  (\bibinfo{year}{1971}).

\bibitem[{\citenamefont{Olver et~al.}(2010)\citenamefont{Olver, Lozier,
  Boisvert, and Clark}}]{olver2010nist}
\bibinfo{author}{\bibfnamefont{F.~W.} \bibnamefont{Olver}},
  \bibinfo{author}{\bibfnamefont{D.~W.} \bibnamefont{Lozier}},
  \bibinfo{author}{\bibfnamefont{R.~F.} \bibnamefont{Boisvert}},
  \bibnamefont{and} \bibinfo{author}{\bibfnamefont{C.~W.} \bibnamefont{Clark}},
  \emph{\bibinfo{title}{NIST handbook of mathematical functions hardback and
  CD-ROM}} (\bibinfo{publisher}{Cambridge University Press},
  \bibinfo{year}{2010}).

\end{thebibliography}
\end{document}